\documentclass{svjour3}

\usepackage{comment}
\usepackage[usenames,dvipsnames]{color}
\usepackage{epsfig}
\usepackage{xspace}
\usepackage{color}

\usepackage{amsmath}
\usepackage{amssymb}
\usepackage{pifont}
\usepackage{algorithm2e}  
\usepackage{lscape}
\usepackage{subfigure}
\usepackage{makeidx}

\usepackage{graphicx}     

\usepackage{array}

\newcommand{\cor}{{\rm \mathcal{C}}}
\newcommand{\prob}{{\rm \textup{Pr}}}
\newcommand{\explored}{{\rm Explored}}
\newcommand{\Count}{{\rm count}}

\newcommand{\false}{{\rm false}}
\newcommand{\true}{{\rm true}}
\newcommand{\rakpath}{{\rm RA\hbox{-}\kappa path}}

\newcommand{\squishlist}{
 \begin{list}{$\bullet$}
  { \setlength{\itemsep}{0pt}
     \setlength{\parsep}{3pt}
     \setlength{\topsep}{3pt}
     \setlength{\partopsep}{0pt}
     \setlength{\leftmargin}{1.5em}
     \setlength{\labelwidth}{1em}
     \setlength{\labelsep}{0.5em} } }

\newcommand{\squishlisttwo}{
 \begin{list}{$\bullet$}
  { \setlength{\itemsep}{0pt}
     \setlength{\parsep}{0pt}
    \setlength{\topsep}{0pt}
    \setlength{\partopsep}{0pt}
    \setlength{\leftmargin}{2em}
    \setlength{\labelwidth}{1.5em}
    \setlength{\labelsep}{0.5em} } }

\newcommand{\squishend}{
  \end{list}  }

\journalname{Journal of Social Network Analysis \& Mining, 3(4), pp. 899-914 (Dec. 2013)}

\begin{document}

\title{Identifying High Betweenness Centrality Nodes in Large Social Networks}

\author{
Nicolas Kourtellis \and
Tharaka Alahakoon \and
Ramanuja Simha \and
Adriana Iamnitchi \and
Rahul Tripathi
}

\institute{
Nicolas Kourtellis \at
Department of Computer Science and Engineering, University of South Florida, Tampa FL, USA\\
Fax: +1-813-974-5456\\
\email{nkourtel@mail.usf.edu}\\
(Corresponding Author)
\and
Tharaka Alahakoon \at
Permanent Address: 2056 Pinnacle Pointe Drive, Norcross, GA 30071\\
\email{alahakoo@mail.usf.edu}
\and
Ramanuja Simha \at
Permanent Address: Department of Electrical and Computer Engineering, University of Delaware, Newark, DE, USA\\
\email{rsimha@mail.usf.edu}
\and
Adriana Iamnitchi \at
Department of Computer Science and Engineering, University of South Florida, Tampa FL, USA\\
Tel.: +1-813-974-5357\\
Fax: +1-813-974-5456\\
\email{anda@cse.usf.edu}
\and
Rahul Tripathi \at
Department of Computer Science and Engineering, University of South Florida, Tampa FL, USA\\
\email{tripathi@cse.usf.edu}
}

\date{DOI 10.1007/s13278-012-0076-6 \\ Received: 20 January 2012 / Revised: 21 May 2012 / Accepted: 30 May 2012}

\maketitle

\begin{abstract}
This paper proposes an alternative way to identify nodes with high betweenness centrality.
It introduces a new metric, $\kappa$-path centrality, and a randomized algorithm for estimating it, and shows empirically that nodes with high $\kappa$-path centrality have high node betweenness centrality.
The randomized algorithm runs in time $O(\kappa^{3}n^{2-2\alpha}\log n)$ and outputs, for each vertex $v$, 
an estimate of its $\kappa$-path centrality up to additive error of $\pm n^{1/2+ \alpha}$ 
with probability $1-1/n^2$.
Experimental evaluations on real and synthetic social networks show improved accuracy in detecting high betweenness centrality nodes and significantly reduced execution time when compared to existing randomized algorithms. 
\end{abstract}

\textbf{Keywords:}
betweenness centrality, social network analysis, algorithms, experimental evaluation

\clearpage

\section{Introduction}
\label{sec:introduction}

Social network analysis tools have been used in various fields such as
physics, biology, genomics, anthropology, economics, organizational
studies, psychology, and IT. The recent phenomenal growth of online
social networks exacerbates the need for such tools that are scalable
for applications in military, government, and for commercial purposes, to name only a few. Some of the 
relevant network metrics are local, such as degree centrality, while others
capture global structural properties of the graph, such as the \textit{betweenness centrality}.
This important global graph metric is a centrality index
that quantifies the importance of a node or an edge as a function of the number
of shortest paths that traverse it.

Node betweenness centrality is relevant to problems such as
identifying important nodes that control flows of information between separate parts of the network and identifying causal nodes to influence other entities behavior, such as genes in genomics or customers in marketing studies.
Betweenness centrality has been used to analyze social networks~\cite{kahng-et-al:j:betcen-corr,lil-edl-ama-sta-abe:j:sex,ort-hoy-lop:j:academic,said-et-al:j:socnet-author} and protein networks~\cite{jeo-mas-bar-olt:j:protein},
to identify significant nodes in wireless ad hoc networks~\cite{maglaras11adhoc},
to study the importance and activity of nodes in mobile phone call networks~\cite{catanese12forensic} and interaction patterns of players on massively multiplayer online games~\cite{ang11mmog},
to study online expertise sharing communities such as physicians~\cite{hua12sna-physicians},
to identify and analyze linking behavior of key bloggers in dynamic networks of blog posts~\cite{macskassy11blogs}
and to measure network traffic in communication networks~\cite{sin-gup:j:congestion}.

Node betweenness centrality, however, is computationally expensive.
The best known algorithm for computing exact betweenness centrality of all vertices 
is Brandes' algorithm~\cite{bra:j:betweenness}, which takes time 
$O(nm)$ on unweighted graphs and $O(nm+n^2\log n)$ 
on weighted graphs.  Some randomized algorithms for estimating betweenness 
centrality have been proposed 
in the literature~\cite{bad-kin-mad-mih:c:approx-betweenness,bra-pic:j:central-estimate,jac-kos-leh-pee-pod:b:algo-for-centrality}, but the accuracy of these randomized algorithms decreases and the execution time increases considerably with the increase in the network
size.
Variants of betweenness centrality, such as flow
betweenness~\cite{fre-bor-whi:j:flow-betweenness} and random-walk
betweenness~\cite{new:j:random-walk}, take computation time at least
of the order $nm$. Thus, existing approaches for exactly computing or even estimating node betweenness
centrality are infeasible for networks with millions of nodes and edges.

We introduce a new approach for identifying highly influential nodes
based on their betweenness centrality score, according to the
following observations. First, we observe that the exact value of the
betweenness centrality 
is irrelevant for many applications: it is the relative ``importance'' of nodes (as measured
by betweenness centrality) that matters. Second, we observe that for the vast majority
of applications, it is sufficient to identify categories of nodes of
similar importance: thus, identifying the top 1\% most 
important nodes is significantly more relevant than precisely ordering
the nodes based on their relative betweenness centrality. Third, we observe 
that distant nodes in (social) networks are unlikely to influence each
other~\cite{bor-eve:j:graph-central,friedkin:j:hor-obser}. Finally, we use the observation that influence may not be
restricted to shortest paths~\cite{ste-zel:j:centrality}. Capturing
these observations, we introduce a new distance-based centrality  
index called $\kappa$-\emph{path centrality}, present a randomized
algorithm for estimating it, provide a complexity and accuracy
analysis of this algorithm, 
and show empirically that nodes with high $\kappa$-path centrality
have high betweenness centrality.   

The contributions of this paper are as follows. First, we introduce a
new node centrality measure, $\kappa$-path centrality, that
is intuitively more appropriate for very large social networks because
it limits graph exploration to a useful neighborhood of $\kappa$ social hops around each node.
The supporting intuition is twofold: first, in social
networks, distant nodes are unlikely to influence each other, and
thus the (long) shortest path that connects them is irrelevant in
practice. Second, shortest paths are not always the choice for information 
transmission, as information may travel on less optimal paths. 

Second, we introduce and evaluate a randomized algorithm that estimates the $\kappa$-path centrality index
for all nodes in a network of size $n$, up to an additive error of at most $n^{1/2 + \alpha}$ with 
probability at least $1 - 1/n^2$ in time $O(\kappa^3 n^{2 - 2 \alpha} \log n)$, where 
$\alpha \in [-1/2, 1/2]$ controls the tradeoff between accuracy and
computation time. 

Third, we demonstrate empirically on a set of real and synthetic social
networks that nodes with high $\kappa$-path centrality have high
betweenness centrality. Moreover, we show that the
running time of our randomized algorithm  
for estimating $\kappa$-path centrality is orders of magnitude lower than 
the runtime of the best known algorithms for computing exact or
approximate betweenness centrality, while maintaining higher accuracy,
especially in very large networks. 
This paper extends our previous work presented in~\cite{alahakoon11kpath-sns} by comparing the k-path measure with other betweenness variants found in the literature, by providing a complexity analysis of the proposed randomized algorithm and by including a more thorough empirical evaluation of the algorithm on 8 new real networks.

In the remaining of the paper we briefly overview the main results in
computing betweenness centrality in
Section~\ref{subsec:betweenness}. We introduce the
$\kappa$-path centrality index and present and analyze the complexity
of the randomized algorithm for
computing it in Section~\ref{sec:kpath}. Section~\ref{sec:evaluation} presents our
experimental results, comparison with Brandes' algorithm, and
two randomized algorithms for estimating betweenness centrality. We
conclude in Section~\ref{sec:summary}.

\section{Node Betweenness Centrality}
\label{subsec:betweenness}
Node betweenness centrality is a global centrality index that quantifies
how much a vertex controls the information flow between all pairs of vertices in a graph. In this
section, we review the formal definition of node betweenness centrality and
briefly overview algorithms used in the experimental evaluation that compute exact and approximate betweenness of all vertices in a graph.

\subsection{Definition and Notations}
Let $G = (V,E)$ be any (directed or undirected) graph, described by the set of vertices $V$ and set of
edges $E$. The number of vertices (edges) in $G$ is denoted by $n$ (respectively, $m$). Let $W$ be 
a non-negative weight function on the edges of $G$, where we assume without loss of generality that each edge 
$e$ of $G$ has $W(e) = 1$ if $G$ is unweighted. 
We define the \emph{length} of any path $\rho$ in $G$ as the sum of weights of edges in $\rho$. A \emph{shortest path} from $s$ to $t$ in $G$ 
is a path of minimum length, and we denote this length by $d_G(s, t)$. Let  
$P_{s}(t)$ denote the \emph{set of predecessors} of a vertex $t$ on shortest paths from $s$ to $t$ in $G$. 
Let $\sigma_{st}$ denote the \emph{number of shortest paths} from $s$ to $t$ in $G$ and, for any $v \in V$, 
let $\sigma_{st}(v)$ denote the number of shortest paths from $s$ to $t$ in $G$ that go through $v$.
Note that $d_G(s, s) = 0$, $\sigma_{ss} = 1$, and $\sigma_{st}(v) = 0$ if $v \in \{s, t\}$ or if 
$v$ does not lie on any shortest path from $s$ to $t$. 

The \emph{betweenness centrality} index of a vertex $v$ is the summation over all pairs of end vertices 
of the fractional count of shortest paths going through $v$. 

\begin{definition}
\label{def:Betweenness Centrality}
{\bf(Betweenness Centrality~\cite{ant:t:rush-digraphs,fre:j:betweenness})}
For every vertex $v$$\in$$V$ of a weighted graph $G(V, E)$,
the \textup{betweenness centrality} $\cor_{B}(v)$ of $v$ is defined by
\begin{equation}
\cor_{B}(v) = \displaystyle\sum_{s \neq v} \displaystyle\sum_{t \neq v, s} 
\frac{\sigma_{st}(v)}{\sigma_{st}}
\end{equation}
\end{definition}

\subsection{Brandes' Algorithm}  
Brandes' algorithm~\cite{bra:j:betweenness} for computing betweenness centrality defines the notion of 
the \emph{dependency score} of any source vertex $s$ on another vertex $v$ as $\delta_{s \star}(v) = \sum_{t \neq s, v} \frac{\sigma_{st}(v)}{\sigma_{st}}$. 
Notice that the betweenness centrality $\cor_{B}(v)$ of any vertex $v$ can be expressed in terms of dependency scores as
$\cor_{B}(v) = \sum_{s \neq v} \delta_{s\star}(v)$.
The following recurrence relation on $\delta_{s\star}(v)$ is significant to Brandes' algorithm:
\begin{equation}
\delta_{s\star}(v) = \sum_{u:v \in P_{s}(u)}\frac{\sigma_{sv}}{\sigma_{su}}(1+\delta_{s\star}(u))
\end{equation}
The algorithm takes as input a graph $G$=$(V,E)$ and an array $W$ of edge weights and outputs the betweenness centrality $\cor_{B}[v]$ of every $v \in V$. 
The running time of Brandes' algorithm on weighted graphs is  
$\mathcal{O}(nm+n^2\log n)$ if the min-priority queue $Q$ is implemented by a \emph{Fibonacci heap}.  
Using BFS instead of Dijkstra's algorithm when the input graph is unweighted, the running time of 
Brandes'  algorithm reduces to $\mathcal{O}(nm)$. 
The space complexity of Brandes' algorithm on both weighted and unweighted graphs is $O(m+n)$. 

\subsection{RA-Brandes Algorithm} 
\label{sec:RA-Brandes}
Adapting the technique of Eppstein and Wang~\cite{epp-wan:j:approx-centrality} 
for estimating the closeness centrality, Jacob et al.~\cite{jac-kos-leh-pee-pod:b:algo-for-centrality} and, independently, Brandes and Pich~\cite{bra-pic:j:central-estimate} 
proposed a randomized approximation algorithm for estimating the betweenness centrality of all vertices in any given graph. 
This algorithm, which we refer to as \emph{Randomized-Approximate Brandes} or in short  \emph{RA-Brandes},  
is different from Brandes' algorithm in only one main respect: Brandes' algorithm considers dependency scores 
$\delta_{s\star}(\cdot)$ of all $n$ start vertices (also called pivots) $s$, whereas RA-Brandes considers dependency scores 
of only a multiset ${\cal S}$ of $\Theta((\log n)/\epsilon^{2})$ pivots. 
The multiset ${\cal S}$ of pivots is selected by choosing vertices uniformly at random with replacement. 
The estimated betweenness centrality $\widehat{\cor}_{B}[v]$ of any vertex $v$ is then defined 
as the scaled average of these scores:
\begin{equation}
\widehat{\cor}_{B}[v] =  \frac{n}{|{\cal S}|}  \sum_{s \in {\cal S}} \delta_{s\star}(v)
\end{equation}
The running time of RA-Brandes on unweighted graphs is $O(\frac{\log n}{\epsilon^{2}}(m + n))$, 
and on weighted graphs is $O(\frac{\log n}{\epsilon^{2}}(m + n\log n))$ if the min-priority queue $Q$ is 
implemented by a Fibonacci heap.
Its space usage on both weighted and unweighted graphs is $O(m+n)$. 
The algorithm guarantees computing, for each vertex $v$, an approximation $\widehat{\cor}_{B}[v]$ that is 
within $\cor_{B}[v] \pm  \epsilon n(n-1)$ with high probability $1-1/n^{\Omega(1)}$. 

\subsection{AS-Brandes Algorithm}
\label{sec:AS-Brandes}
Bader et al.~\cite{bad-kin-mad-mih:c:approx-betweenness} proposed a randomized algorithm for estimating the betweenness centrality 
of all vertices in any given graph. Their algorithm is based on the \emph{adaptive sampling} technique of Lipton and Naughton~\cite{lip-nau:c:transitive} 
used in an algorithm for estimating the size of the transitive closure of a directed graph. The adaptive sampling technique requires selecting a multiset 
of start vertices by sampling vertices adaptively in the sense that the number of vertices chosen varies with the information gained from each sample. 
To precisely bound the running time, this algorithm terminates when the number of samples reaches a predetermined cut-off $T$ supplied to the algorithm.
Because of its similarity to Brandes' algorithm and application of adaptive sampling technique, we refer to this algorithm as \emph{Adaptive-Sampling Brandes} or in short \emph{AS-Brandes}. 

The algorithm AS-Brandes considers dependency scores of only a multiset ${\cal S}$ of at most $T$ pivots. 
It estimates betweenness centrality of any vertex $v$ by noting how fast the sum of dependency scores 
for $v$ reach a threshold $cn$, where $c \geq 2$ is supplied to the algorithm. To this end, for each vertex 
$v$, the algorithm maintains a running sum $RS[v]$ of dependency scores $\delta_{s\star}(v)$ for pivots $s$ and 
it records in a variable $k[v]$, the number of pivots used for $v$ until $RS[v]$ becomes greater than $cn$; 
$k[v]$ is set to $T$ if $RS[v]$ never exceeds $cn$.  The estimated betweenness centrality $\widehat{\cor}_{B}[v]$ 
of any vertex $v$ is then defined as the scaled average of these scores over $k[v]$ samples: 
\begin{equation}
\widehat{\cor}_{B}[v] = n \frac{RS[v]}{k[v]}
\end{equation}
Since AS-Brandes considers only $T$ pivots while Brandes' algorithm considers all $n$ pivots, 
AS-Brandes should be roughly $\Omega(n/T)$ times faster than Brandes' algorithm. 
The space usage of AS-Brandes on both weighted and unweighted graphs is $O(m+n)$. 
The algorithm guarantees that, 
for $0 < \epsilon < 0.5$, if the betweenness centrality $\cor_{B}[v]$ of a vertex $v$ is at least $n^2/t$ for some constant 
$t \geq 1$, then with probability at least $1-2\epsilon$, its estimated betweenness centrality $\widehat{\cor}_{B}[v]$ 
is within $(1 \pm 1/\epsilon)\cdot \cor_{B}[v]$ using $\epsilon t$ pivots.

\section{K-Path Centrality}
\label{sec:kpath}

As introduced in~\cite{new:j:random-walk}, the random-walk betweenness
centrality is based on  
the traversal of the network with absorbing random walks.  
Assume the traversal of a message (e.g., news or rumor) originating
from some source $s$ over a network and intending to finally reach
some destination $t$ in the network along a path,  
and assume that each node in the network has only its own local view
(i.e., has information only of its outgoing neighbors). 
Thus, when the message is at a current node $v$, the node $v$ 
forwards the message based on its local view to one of its outgoing
neighbors chosen uniformly at random.  
The message continues to travel in this manner until it reaches the
destination node $t$, and then stops. 

The notion of $\kappa$-path centrality is based on a similar
assumption regarding the random traversal of a message from  a source
$s$. However, we make two further assumptions in order to reduce the
computation time without deviating much  from the above random walk
model. First, we consider message traversals along simple paths only,
i.e., paths in which vertices do not repeat. As non-simple paths do
not correspond  to the intuitive notion of ideal message traversals in
a social network, their consideration in the computation of centrality
indices is a noisy factor. To discount non-simple paths, we assume
that each intermediate node $v$ on a partially traversed  path
forwards the message to a neighbor chosen randomly, with probability
inversely proportional to edge weights, from the current set of
unvisited neighbors; the message traversal is assumed to stop if all
the outgoing neighbors of the current node $v$ already appear in the
path up to $v$.  Although choosing a random neighbor in this manner at
each step requires the premise that the message carries the history  
of the path traversed so far, this premise is needed to express the
average contribution of any simple path in the overall information flow and to efficiently
simulate such random simple paths. 
Second, we assume that the message traversals are only along paths of at most $\kappa$ edges, where $\kappa$ is a parameter dependent on the network. 
It has been found in many studies on social networks that message traversals typically take paths containing few edges~\cite{friedkin:j:hor-obser}, and so this seems to be a reasonable assumption in the context of social networks.
Based on these assumptions, we define $\kappa$-path centrality: 

\begin{definition}
\label{def:kpath-informal} 
{\bf \boldmath ($\kappa$-path centrality)} For every vertex $v$ of a graph $G = (V,E)$, the  $\kappa$-path centrality $\cor_{k}(v)$ 
of $v$ is defined as the sum, over all possible source nodes $s$,  of the probability that a message originating from $s$ goes 
through $v$, assuming that the message traversals are only along random simple paths of at most $\kappa$ edges.
\end{definition}

\subsection{Formal Analysis of K-Path Centrality}
\label{subsec:K-path-Centrality}
\noindent
Consider an arbitrary simple path $\rho_{s,\ell}$  with start vertex $s$ and having $\ell \leq \kappa$ edges, where $\kappa$ is the 
value of parameter $K$ in $K$-path centrality.  
Let $s$, $u_1$, $u_2$, $\ldots$, $u_{\ell-1}$, $u_{\ell}$ denote the vertices in the order they appear in $\rho_{s,\ell}$ and $s=u_0$ for convenience. For every $0 \leq i \leq \ell$, 
let $(s, u_1, \ldots, u_{i-1}, u_{i})$ denote $\rho_{s,i}$, the subpath from $s$ to $u_i$, and let $\prob[\rho_{s,i}]$ denote the probability 
that a message originating from $s$ traversed through the path $\rho_{s,i}$. The probability $\prob[\rho_{s,\ell}]$, as shown below, is equal to 
the product of individual probabilities associated with the random transitions of the message between successive nodes of $\rho_{s,\ell}$. 
The exact expression of $\prob[\rho_{s,\ell}]$ depends on whether the graph is weighted or unweighted; so we consider these two cases separately.

Consider the case of an unweighted, directed graph in which $\rho_{s,\ell}$ is a simple path from $s$ to $u_{\ell}$. 
For every $0 \leq i \leq \ell$, let 
$N(u_i)$ denote the set of outgoing neighbors of $u_i$. The expression for $\prob[\rho_{s,\ell}]$ is given by the following 
recurrence relation:

\begin{equation} 
\prob[\rho_{s,i}] =
\begin{cases} 
\prob[\rho_{s,i-1}] \times \prob[\textnormal{edge $(u_{i-1}, u_i)$ is chosen given $\rho_{s,i-1}$}] & \mbox{if $i \geq 2$} \\
\frac{1}{|N(s)|} & \mbox{if $i = 1$}
\end{cases}
\end{equation}

Here, $\prob[\textnormal{edge $(u_{i-1}, u_i)$ is chosen given $\rho_{s,i-1}$}]$ denotes the conditional probability that 
the message is forwarded from $u_{i-1}$ to $u_{i}$, given that the path 
traversed up to $u_{i-1}$ is $\rho_{s,i-1}$. This probability is equal to $1/|N(u_{i-1}) - \{s, u_1, u_2, \ldots, u_{i-2}\}|$, 
since, by our assumption, each node $u_i$ forwards the message to a node chosen uniformly at random from the unvisited 
neighbors of $u_i$. The above recurrence relation easily leads to the following solution: 
\begin{equation}
\prob[\rho_{s,\ell}] = \prod_{i=1}^{\ell} \frac{1}{|N(u_{i-1}) - \{s, u_1, u_2, \ldots, u_{i-2}\}|}. \label{eq:prho-unweighted}
\end{equation}
Notice from the above expression that the larger the outdegree of a node is, the smaller the probability of the message being 
forwarded through a specific edge is. This observation corresponds to the intuition that, if the intermediates nodes of a path 
have a high outdegree, then it is less likely for a message from the source to take that path in its entirety. 

Next consider the case of a weighted, directed graph in which $\rho_{s,\ell}$ is a simple path from $s$ to $u_{\ell}$. In this case, each edge 
$(u_{i-1}, u_{i})$ in $\rho_{s,\ell}$ has a weight $W(u_{i-1}, u_{i})$. Intuitively, the weight of the edge $(u_{i-1}, u_{i})$ quantifies how easily any 
information from $u_{i-1}$ can pass to $u_{i}$: the smaller the weight of an edge is, the more accessible the endpoint of the edge is. 
Thus, it is more likely for a message to be forwarded on to a lower weight edge than to be forwarded on to a higher weight edge from any node. 
This intuition suggests the following analog of Eq.~\eqref{eq:prho-unweighted} for the case of weighted graphs: 
\begin{equation}
\prob[\rho_{s,\ell}] = \prod_{i=1}^{\ell} \frac{1/W(u_{i-1}, u_{i})}{\sum_{v \in N(u_{i-1}) - \{s, u_1, u_2, \ldots, u_{i-2}\}} 1/W(u_{i-1},v)}.
\label{eq:prho-weighted}
\end{equation}
Here, the conditional probability that the message is forwarded from $u_{i-1}$ to $u_i$, given that the path traversed up to $u_{i-1}$ is 
$\rho_{s,i-1}$, is given by the expression within the product symbol. In this expression, the numerator $1/W(u_{i-1}, u_{i})$ corresponds to the 
intuition that the probability of the message traversing the edge $(u_{i-1}, u_i)$ is inversely proportional to the weight of this edge and the 
denominator is only a normalization factor so that the probabilities sum to one. 

With the above expression for $\prob[\rho_{s,\ell}]$, we now formalize the notion of $\kappa$-path centrality. 
For any simple path $\rho_{s,\ell}$ originating from $s$ and any $v \neq s$, let 
\begin{equation}
\chi[v \in \rho_s] =
\begin{cases}
 1 & \mbox{\text{if v lies on $\rho_s$, and}} \\
 0 & \mbox{\text{otherwise}}
\end{cases}
\end{equation} 

Then, the probability that the message originating from $s$ goes through any vertex $v$ as per our assumptions is given by 
\begin{equation} 
\sum_{1 \leq \ell \leq \kappa} \sum_{\rho_{s,\ell}: |\rho_{s,\ell}| = \ell} \chi[v \in \rho_{s,\ell}] \cdot  \prob[\rho_{s,\ell}].
\end{equation} 
The first summation is over the edge counts $\ell$ of any simple path and the second summation is over 
all simple paths $\rho_{s,\ell}$ whose edge count is exactly $\ell$. In these summations, the contribution 
$\prob[\rho_{s,\ell}]$ of any simple path $\rho_{s,\ell}$ is included if and only if $v$ lies on $\rho_{s,\ell}$, as indicated by the 
expression $\chi[v \in \rho_{s,\ell}] \cdot  \prob[\rho_{s,\ell}]$.  Thus, we get an alternative formulation 
of $\kappa$-path centrality.

\begin{proposition}
\label{def:kpath}
{\bf \boldmath ($\kappa$-path centrality)} 
For every vertex $v$ of graph $G$=$(V, E)$, the \textup{$\kappa$-path centrality} $\cor_{k}(v)$ of $v$ is given by 
\begin{equation}
\cor_{k}(v) = \displaystyle \sum_{s \neq v} \text{  } \displaystyle \sum_{1 \leq \ell \leq k} \text{  } \sum_{\rho_{s,\ell}: |\rho_{s,\ell}| = \ell} \chi[v \in \rho_{s,\ell}] \cdot  \prob[\rho_{s,\ell}], 
\end{equation}
where  $\prob[\rho_{s,\ell}]$ is described by Eq.~\eqref{eq:prho-unweighted} if $G$ is unweighted and by Eq.~\eqref{eq:prho-weighted} if $G$ is weighted. 
\end{proposition}

\subsection{Comparison With Variants of Betweenness}\label{sec:comparison}

The notion of $\kappa$-path centrality contrasts with other variants of 
betweenness (e.g., $k$-betweenness, random-walk betweenness and flow betweenness) in 
definitions, assumptions, as well as algorithmic complexity. 

\textbf{$k$-betweenness or bounded-distance betweenness:}
Betweenness centrality considers contributions from all shortest paths irrespective of their length. 
Borgatti and Everett~\cite{bor-eve:j:graph-central} suggested the idea of limiting the length of shortest paths in the definition of betweenness centrality, 
as they argued that long paths are seldom used for propagation of influence in some networks. They defined $k$-betweenness centrality as an 
index in which, for each vertex $v$, its centrality (similar to the case of betweenness) is the sum of dependency scores $\delta_{s\star}(v)$ of all start vertices $s$ on $v$, but the dependency scores account for only those shortest paths that are of length at most $k$ (as opposed to the case of betweenness in which contributions from all shortest paths are considered).
Later, Brandes~\cite{bra:j:spath-betweenness} redefined this measure as bounded-distance betweenness centrality.
For every vertex $v \in V$ of a graph $G = (V, E)$, the $k$-\emph{betweenness centrality}~\cite{bor-eve:j:graph-central} $\cor_{B(k)}(v)$ of $v$ is defined as $ \cor_{B(k)}(v) = \sum_{s, t \in V: d_G(s, t) \leq k} \frac{\sigma_{st}(v)}{\sigma_{st}} $.
The $k$-betweenness centrality of all vertices of a graph can be computed using Brandes' algorithm where we stop the underlying single-source shortest path search when a vertex of distance $k$ from the source is reached.
In traversing the graph from every (source) vertex to all other vertices, the single-source shortest path search breaks on reaching the first vertex that is at distance at least $k$ from the source.
In the worst case, if the shortest path distances from every vertex to all other vertices are no more than $k$, then the algorithmic complexity will be identical to Brandes' algorithm.

\textbf{Random-walk betweenness:} Introduced by Newman~\cite{new:j:random-walk}, it assumes that message transmission 
between any two individuals $s$ and $t$ in a social network follows a random path. It models the path the message takes 
as an absorbing random walk from $s$ to $t$. The net flow of this random walk on an edge 
$\{x,y\}$ is defined as the absolute difference between the probability that the walk goes from $x$ to $y$ and the probability that it goes 
from $y$ to $x$. The net flow of the random walk through vertex $x$ is defined as one-half of the sum of the net flows on the edges incident 
to $x$. The net flow (along an edge or a vertex) is defined in this way so as to discount the possibility that a random walk repeats a vertex 
or an edge multiple times. The random-walk betweenness of a vertex $v$ is the expected net flow of a random walk from source $s$ to 
destination $t$ through $v$, where the expectation is over all possible pairs $(s,t)$. The best known algorithm for exactly computing random-walk 
betweenness of all vertices takes time $O(I(n-1) + mn\log n)$, where $I(n) = O(n^3)$ is the time for computing the inverse of an 
$n \times n$-matrix~\cite{bra-fle:c:central}. 

\textbf{Flow betweenness:} Introduced by Freeman et al.~\cite{fre-bor-whi:j:flow-betweenness}, it models any directed network as a flow network where 
edges represent pipes that can carry up to unit amount of flow. The model assumes a flow to be generated at a source node $s$, 
transmitted across edges, and absorbed at a sink node $t$. The value of the flow is defined as the total amount of flow generated at $s$ 
and the amount of flow through any vertex $x$ is the total amount of flow leaving $x$. 
This notion requires determining the quantity of the flow through a particular vertex $v$ assuming that the flow transmitting 
from $s$ to $t$ has the maximum possible value. (In case this quantity is not unique because more than one solutions exist for the 
$st$-maximum flow problem, then we seek for the maximum flow through $v$ over all possible solutions.) 
The flow betweenness of a vertex $v$ is defined as the average of this quantity over all possible source-sink pairs $(s,t)$. The flow betweenness 
of all vertices can be exactly computed in time $O(m^2n)$ as reported in~\cite{new:j:random-walk}. 

\subsection{Estimating K-Path Centrality with a Randomized Approximation Algorithm}
\label{sec:randomized-algo}

We present a randomized approximation algorithm for estimating the $\kappa$-path centrality of all vertices in any graph.  
The algorithm takes as input a graph $G$=$(V,E)$, a non-negative weight function $W$ on the edges of $G$, 
and parameters $\alpha$$\in$$[-1/2, 1/2]$ and integer $\kappa$=$f(m,n)$, and runs in time $O(\kappa^{3}n^{2-2\alpha} \ln n)$. 
For each vertex $v$, it outputs an estimate of $\cor_{\kappa}(v)$ up to an additive error of $\pm n^{1/2+\alpha}$ with probability at least $1-1/n^2$. 
We refer to this algorithm as \emph{Randomized-Approximate $\kappa$path} or in short \emph{RA-$\kappa$path}.

The algorithm, shown in Algorithm~\ref{algo:approx-kpath}, performs $T = 2\kappa^{2}n^{1-2\alpha}\ln n$ iterations (the expression for $T$ comes from the analysis of the algorithm, shown next). In each iteration, a start vertex $s \in V$ 
and a walk length $\ell \in [1, \kappa]$ are chosen uniformly at random.
In every iteration, a random walk consisting of 
$\ell$ edges from $s$ is performed, which essentially simulates a message traversal from $s$ in $G$ using the assumption made in Definition~\ref{def:kpath-informal}.
The number of times any vertex $v$ is visited over all the random walks is recorded in a variable $\Count[v]$. The estimated $\kappa$-path centrality  $\widehat{\cor}_{\kappa}[v]$ of any vertex $v$ is then defined as the scaled average of the times $v$ is visited over $T$ walks: $\widehat{\cor}_{\kappa}[v] = \kappa n \cdot \frac{\Count[v]}{T}$. 

\clearpage
\scriptsize{
\begin{algorithm}[tpbh]
\begin{tabbing}
\hspace*{0.5cm}\=\hspace{0.5cm}\=\hspace{0.5cm}\=\hspace{0.5cm}\=\hspace{0.5cm}\= \kill
\texttt{Input: Graph $G = (V,E)$, Array $W$ of edge weights,}\\
\> \> \texttt{$\alpha \in [-1/2, 1/2]$ and integer $\kappa$} \\
\texttt{Output: Array $\widehat{\cor}_{\kappa}$ of $\kappa$-path centrality estimates} \\
\texttt{\textbf{begin}} \\
\> \texttt{\textbf{foreach} $v \in V$ \textbf{do}} \\
\> \> \texttt{$\Count[v] \leftarrow 0$;  $\explored[v] \leftarrow \false$;} \\
\> \texttt{\textbf{end}} \\
\> \texttt{/* $S$ is a stack and $n = |V|$ */} \\
\> \texttt{$T \leftarrow 2\kappa^{2}n^{1-2\alpha}\ln n$; $S \leftarrow \emptyset$;} \\
\> \texttt{\textbf{for} $i \leftarrow 1$ \textbf{to} $T$ \textbf{do}} \\
\> \> \texttt{/* simulate a message traversal from $s$ over $\ell$ edges */} \\
\> \> \texttt{$s$ $\leftarrow$ a vertex chosen uniformly at random from $V$;} \\
\> \> \texttt{$\ell \leftarrow$ an integer chosen uniformly at random from $[1, \kappa]$;} \\
\> \> \texttt{$\explored[s] \leftarrow$ true; push $s$ to $S$;} \texttt{$j \leftarrow 1$;}\\  
\> \> \texttt{\textbf{while} $(j \leq \ell$ and $\exists (s, u) \in E$ s.t.~$!\explored[u])$ \textbf{do}} \\
\> \> \> \texttt{$v \leftarrow$ a vertex chosen randomly from $\{u \mid (s,u) \in E$} \\
\> \> \> \> \texttt{and $!\explored[u]\}$ with probability} \\
\> \> \> \> \texttt{proportional to $1/W(s,v)$;} \\  
\> \> \> \texttt{$\explored[v] \leftarrow \true$; push $v$ to $S$;} \\  
\> \> \> \texttt{$\Count[v] \leftarrow \Count[v] + 1$;} \\ 
\> \> \> \texttt{$s \leftarrow v$;} \texttt{$j \leftarrow j + 1$;}\\
\> \> \texttt{\textbf{end}} \\
\> \> \texttt{/* reinitialize $\explored[v]$ to false */} \\
\> \> \texttt{\textbf{while} $S$ is nonempty \textbf{do}} \\
\> \> \> \texttt{pop $v \leftarrow S$;} \texttt{$\explored[v] \leftarrow \false$;}\\ 
\> \> \> \texttt{/* if message traversal stops in less than $\ell$ edges,} \\ 
\> \> \> \texttt{reset count values to the old ones */} \\
\> \> \> \texttt{\textbf{if} $(j \leq \ell)$} $\Count[v] \leftarrow \Count[v] - 1$  \\
\> \> \texttt{\textbf{end}} \\
\> \texttt{\textbf{end}} \\
\>  \texttt{\textbf{foreach} $v \in V$ \textbf{do}} \\
\> \> \texttt{$\widehat{\cor}_{\kappa}[v] \leftarrow \kappa n \cdot \frac{\Count[v]}{T}$;} \\
\> \texttt{\textbf{end}} \\
\> \texttt{\textbf{return} $\widehat{\cor}_{\kappa}$;} \\
\texttt{\textbf{end}} \\
\end{tabbing}
\caption{Randomized approximation algorithm for estimating the $\kappa$-path centrality.}
\label{algo:approx-kpath}
\end{algorithm}
}
\normalsize

\begin{theorem}
\label{thm:kpath-guar}
The algorithm $\rakpath$ runs in time $O(\kappa^{3}n^{2-2\alpha}\log n)$ and outputs, for each vertex $v$, 
an estimate $\widehat{\cor}_{\kappa}[v]$ of $\cor_{\kappa}[v]$ up to an additive error of $\pm n^{1/2+\alpha}$ with probability $1-1/n^2$. 
\end{theorem}
{\bf Proof\quad} Fix an arbitrary vertex $v \in V$, real $\alpha \in [-1/2, 1/2]$, and integer $\kappa \geq 1$. 
We define random variables $X_i$, for $1~\leq~i~\leq~T$, corresponding to 
the $T$ iterations as follows
\[ 
X_{i} = 
\begin{cases} 
1   & \text{if the $i$'th random simple path goes through $v$,} \\
0   & \text{otherwise}
\end{cases} 
\]
It is easy to see that when the algorithm terminates, $\Count[v] = \sum_{i=1}^{T} X_{i}$. 
Let us now evaluate the expected value $E[X_i]$ of $X_i$, for any $1 \leq i \leq T$. Since $X_i$ is an indicator random variable, 
we have $E[X_i]$=$\prob[X_i$=$1]$, and, by the definition of $X_i$, $\prob[X_i$=$1]$ equals the probability 
that the $i$'th random simple path goes through $v$.  The algorithm chooses a random start vertex 
$s$ and a random edge count $\ell$$\in$$[1,\kappa]$, where both are distributed uniformly over their respective 
sample sets. Thus, for any vertex $s$ and edge count $\ell$$\in$$[1,\kappa]$, $s$ is chosen as a start vertex and 
$\ell$ is chosen as a edge count with probability $1/\kappa n$. Once $s$ and $\ell$ are fixed, 
then a path $\rho_{s,\ell}$ of $\ell$ edge counts originating from $s$ is traversed with probability $\prob[\rho_{s,\ell}]$, 
described by Eq.~\eqref{eq:prho-unweighted} if $G$ is unweighted and by Eq.~\eqref{eq:prho-weighted} 
if $G$ is weighted. It follows that 
\begin{align}
E[X_i] &= \frac{1}{\kappa n} \sum_{s \neq v} \sum_{1 \leq \ell \leq \kappa} \sum_{\rho_{s,\ell}: |\rho_{s,\ell}| = \ell} \chi[v \in \rho_{s,\ell}] \cdot \prob[\rho_{s,\ell}], \nonumber \\ 
&= \frac{1}{\kappa n} \cor_{\kappa}[v]  \hspace*{1 cm} \text{(by Proposition~\ref{def:kpath})}.   \label{eq:exck}
\end{align} 
Let us define random variables $Y_i$, for $1$$\leq$$i$$\leq$~$T$, as $Y_i$$=$$\kappa$$nX_i$.
Note that $Y_i$s are independent random variables and each $Y_i$ takes value either $0$ or $\kappa n$. 
Also, note that the estimate of $\cor_{\kappa}[v]$ returned by RA-Kpath algorithm is
$\widehat{\cor}_{\kappa}[v]$$=$$\kappa$$n$$\frac{ \Count[v]}{T}$=$\frac{\sum_{i=1}^{T} Y_i}{T}$.
Thus, by linearity of expectation,
\begin{align*}
E[\frac{\sum_{i=1}^{T} Y_i}{T}] &= \frac{\kappa n}{T} E[\sum_{i=1}^{T} X_i] \\
&= \kappa n \cdot E[X_i] \\
&= \cor_{\kappa }(v)  \hspace*{1 cm} \text{(by Eq.~\ref{eq:exck}).} 
\end{align*}
Application of Hoeffding bound\footnote{The Hoeffding bound~\cite{hof:j:bound}, a classical result in probability theory, states the following: 
Let $X_{1}, X_{2}, \ldots, X_{T}$ be independent random variables, such that each $X_i$ ranges over the real interval $[a_i, b_i]$, and 
let $\mu = E\left[\sum_{i=1}^{T} X_{i}/{T}\right]$ denote the expected value of the average of these variables. Then, for every $\xi > 0$, 
$\prob\left[\left|\frac{\sum_{i=1}^{T} X_{i}}{T} - \mu\right| \geq \xi\right] \leq 2e^{-2T^{2}\xi^{2}/\sum_{i=1}^{T} (b_{i} - a_{i})^{2}}$.}
gives 
\begin{align*}
\prob\left[\left|\frac{\sum_{i=1}^{T} Y_{i}}{T} - \cor_{\kappa}(v)\right| \geq \xi\right] &\leq 2e^{-2T^{2}\xi^{2}/(T \kappa^{2}n^{2})} \\ 
&= 2e^{-2T\xi^{2}/(\kappa^{2}n^{2})}.
\end{align*} 
Keeping the error margin $\xi$ to $n^{1/2+\alpha}$ results in
\begin{equation}
\prob[|\widehat{\cor}_{\kappa}[v]-\cor_{\kappa}(v)|\geq \xi]\leq 2e^{-2T/(\kappa^{2}n^{1-2\alpha})}.
\end{equation}
This probability can be made at most $1/n^3$ if $T$$\geq$$2\kappa^{2}n^{1-2\alpha}\ln{n}$. 
Thus, setting $T$ to $2\kappa^{2}n^{1-2\alpha}\ln{n}$ yields, for every vertex $v$, an estimate $\widehat{\cor}_{\kappa}[v]$ 
of $\cor_{\kappa}[v]$ up to an additive error of $\pm n^{1/2+\alpha}$  with probability at least $1 - 1/n^2$.  
In each of the $T$ iterations, the time spent is $O(\kappa n)$.
Therefore, the running time of the algorithm is 
$O(T\kappa n) = O(\kappa^{3}n^{2-2\alpha}\ln n)$. $\Box$

\section{Experimental Evaluation}
\label{sec:evaluation}

\scriptsize{
\begin{table}
\begin{center}
\caption{\small Summary information of the real networks used (d/u: directed/undirected; w/uw: weighted/unweighted).}
\label{tab:real-info}
\footnotesize {
\begin{tabular}{|l|l|l|l|l|l|l|l|} \hline
Real	Networks			&	Nodes	&	Edges		&	d/u, w/uw	&	Ref.						&	Network Type	\\ \hline \hline
Kazaa			&	2424		&	13354		&	u, w		&\cite{Iam-Rip-Fos:c:Small-World}	&	File sharing	\\ \hline
Kazaa (U)			&	2424		&	13354		&	u, uw	&\cite{Iam-Rip-Fos:c:Small-World}	&	File sharing	\\ \hline
SciMet			&	2729		&	10416		&	u, uw	&\cite{Bat-Mrv:w:Pajek-dataset}	& 	Citation   		\\ \hline
Kohonen			&	3772		&	112731		&	u, uw	&\cite{Bat-Mrv:w:Pajek-dataset}	& 	Citation  		\\ \hline
Geom			&	6158		&	11898		&	u, w		&\cite{Bat-Mrv:w:Pajek-dataset}	&	Co-authorship	\\ \hline
Geom (U)			&	6158		&	11898		&	u, uw	&\cite{Bat-Mrv:w:Pajek-dataset}	&	Co-authorship	\\ \hline
CA-AstroPh		&	18772	&	396160		&	u, uw	&\cite{Les:w:Stanford-dataset}		&	Collaboration	\\ \hline
CA-CondMat		&	23133	&	186936		&	u, uw	&\cite{Les:w:Stanford-dataset}		&	Co-authorship	\\ \hline
Cit-HepPh			&	34546	&	421578		&	d, uw	&\cite{Les:w:Stanford-dataset}		&	Citation    		\\ \hline
Email-Enron		&	36692	&	367662		&	u, uw	&\cite{Les:w:Stanford-dataset}		&	Email communication	\\ \hline
Cond-Mat-2005	&	40421	&	175693		&	u, w		&\cite{new:j:struc-sci-coll-networks}	&	Co-authorship	\\ \hline
Cond-Mat-2005 (U)	&	40421	&	175693		&	u, uw	&\cite{new:j:struc-sci-coll-networks}	&	Co-author	ship	\\ \hline
P2P-Gnutella31	&	62586	&	147892		&	d, uw	&\cite{ripeanu02gnutella}			&	File sharing	\\ \hline
Soc-Epinions1		&	75879	&	508837		&	d, uw	&\cite{Les:w:Stanford-dataset}		&	Social		\\ \hline
Soc-Slashdot0902	&	82168	&	948464		&	d, uw	&\cite{Les:w:Stanford-dataset}		&	Social		\\ \hline
\end{tabular}
}
\end{center}
\end{table}
}
\normalsize

In order to assess the performance of the algorithm $\rakpath$, we
compare in Section~\ref{sec:comparison-brandes} its accuracy and running time 
with that of Brandes' algorithm and in Section~\ref{sec:comparison-approx-algs}
with that of the two betweenness centrality approximation algorithms (RA-Brandes and AS-Brandes).
We performed experiments on both real and synthetic social networks.
The real networks selected cover a wide range of application domains and scales (file sharing, citation, co-authorship, collaboration, email communication and social), and are presented in Table~\ref{tab:real-info}.
In order to test the performance of $\rakpath$ on social graphs
that maintain consistent social properties with increase in their size,
we created $10$ independent sets of networks with varying sizes ($1K$, $10K$, $50K$ and $100K$ nodes)
using a synthetic social network generator based on the model in~\cite{scwzzz:c:measurement}.
All experiments were done on a cluster of identical machines with dual core AMD Opteron processors at 2.2 GHz and
4GB RAM.

\subsection{Performance Metrics} 
\label{subsubsec:per-mea}

For evaluating the accuracy of $\kappa$-path centrality in estimating the relative 
importance of a node as per the betweenness centrality index, we chose two accuracy metrics.
The first metric, called $\rakpath$ correlation, is the correlation
between the approximate $\kappa$-path centrality values computed by
$\rakpath$  and the exact betweenness centrality values computed by
Brandes' algorithm, for all users in the graph.
We applied the same approach to measure the accuracy of the other two approximation algorithms, RA-Brandes and AS-Brandes.
We refer to these metrics as RA-Brandes correlation and AS-Brandes correlation, respectively.   

The second accuracy metric captures the ability to identify the $top$--$N\%$ high betweenness centrality nodes.
For this, we measured the percentage of the overlap between the $top$--$N\%$ nodes as returned by
a particular approximation algorithm ($\rakpath$, RA-Brandes, and AS-Brandes) and the $top$--$N\%$ nodes as identified by Brandes' algorithm.
We refer to these metrics as top $N\%$ $\rakpath$, top $N\%$ RA-Brandes, and top $N\%$ AS-Brandes, respectively.

For evaluating the run time performance, we determined the ratio of the
execution time of each of the three approximation algorithms over our
implementation of  Brandes' algorithm. We refer to this performance metric as
speedup, and thus we compare  $\rakpath$ speedup, RA-Brandes speedup, and
AS-Brandes speedup. 

\subsection{Comparison with Brandes' Algorithm}
\label{sec:comparison-brandes}
\begin{figure}[h]
	\vspace*{-1 mm}
	\begin{center}
%

		\includegraphics[scale=0.9]{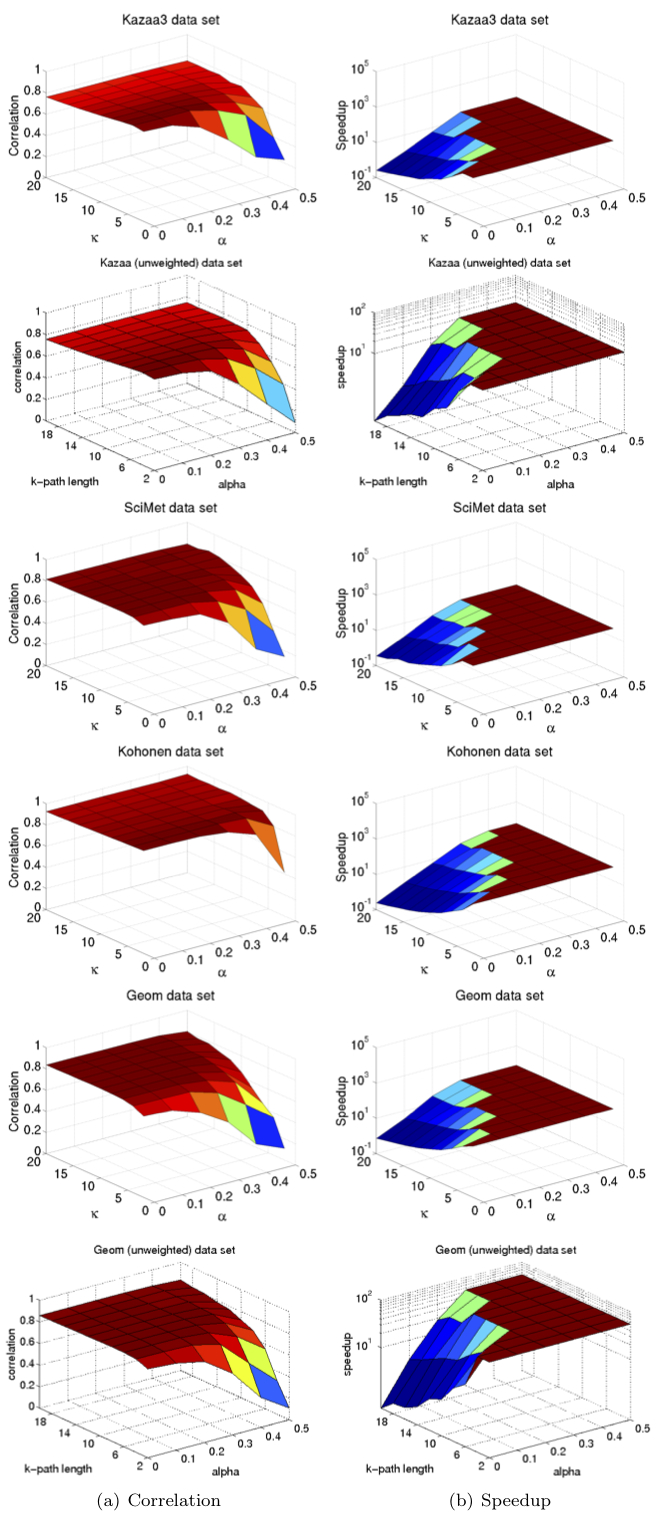}

	\end{center}
	\vspace*{-3 mm}
	\caption{\small RA-$\kappa$path correlation (a) and speedup (b) for the real networks with size below $10K$ nodes.}%
	\label{fig:corr-speedup-10K}
\end{figure}

\begin{figure}[h]
	\vspace*{-1 mm}
	\begin{center}
%

		\includegraphics[scale=0.9]{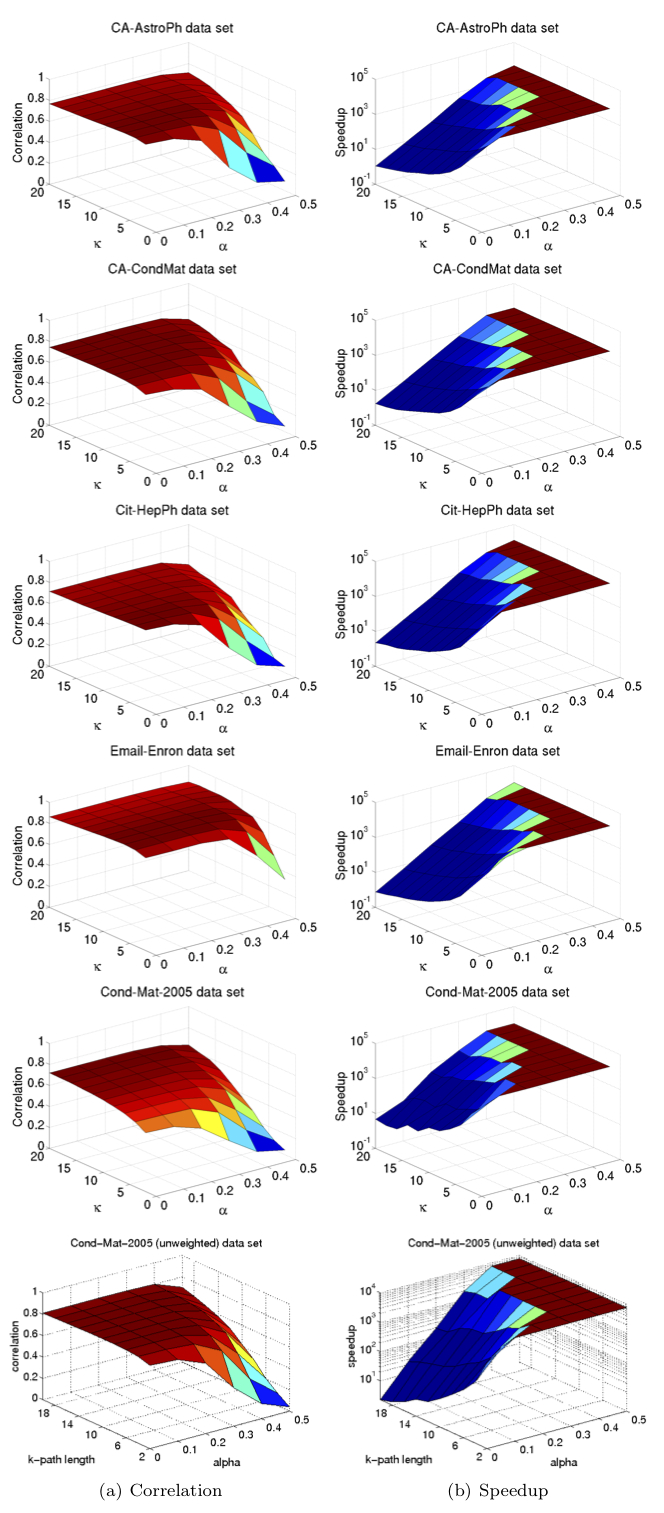}

	\end{center}
	\vspace*{-3 mm}
	\caption{\small RA-$\kappa$path correlation (a) and speedup (b) for the real networks with size between $10K$ and $50K$ nodes.}%
	\label{fig:corr-speedup-50K}
\end{figure}

\begin{figure}[h]
	\vspace*{-1 mm}
	\begin{center}
%

		\includegraphics[scale=0.6]{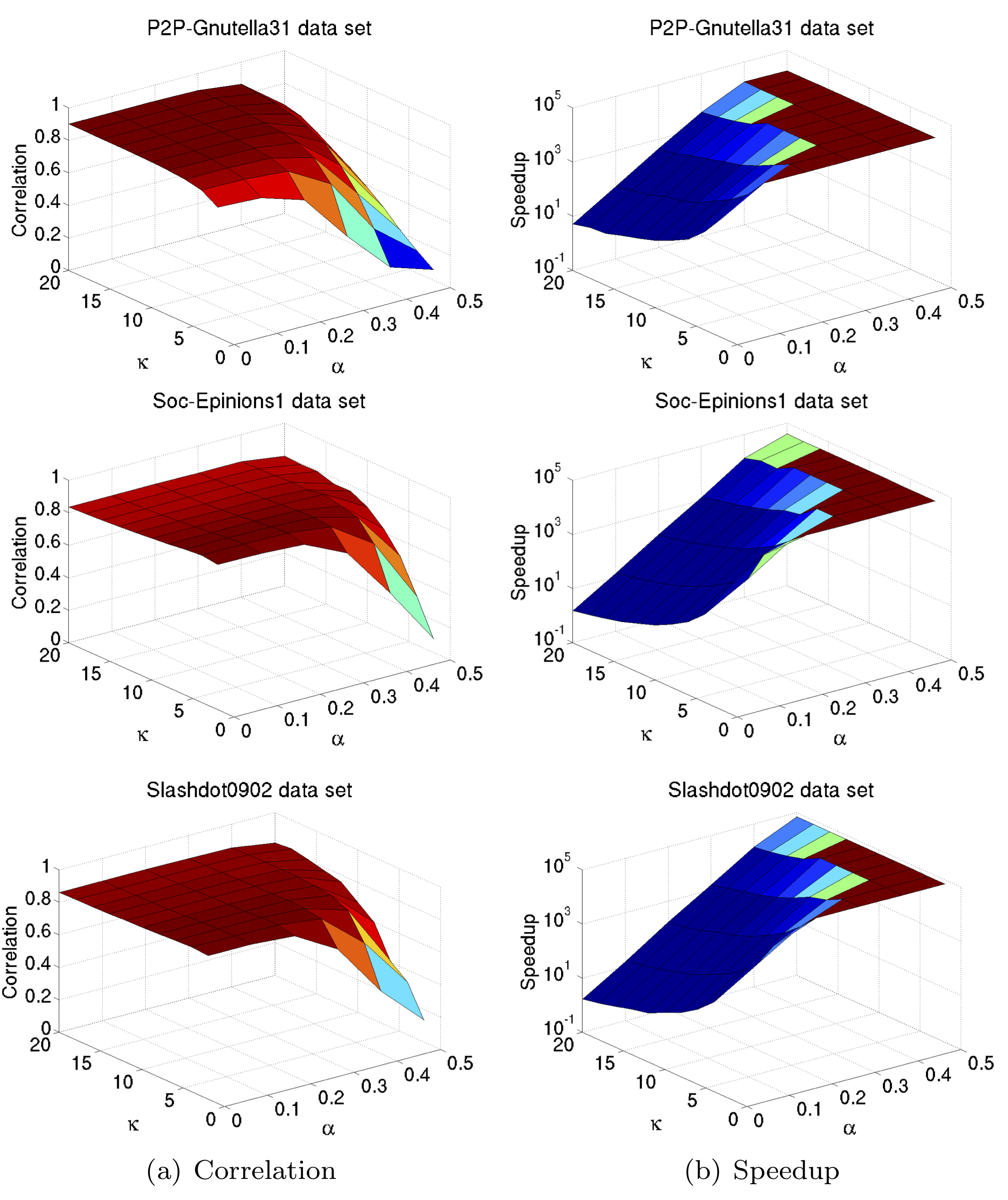}

	\end{center}
	\vspace*{-3 mm}
	\caption{\small RA-$\kappa$path correlation (a) and speedup (b) for the real networks with size between $50K$ and $100K$ nodes.}%
	\label{fig:corr-speedup-100K}
\end{figure}

\begin{figure}[h]
	\vspace*{-1 mm}
	\begin{center}

		\includegraphics[scale=0.6]{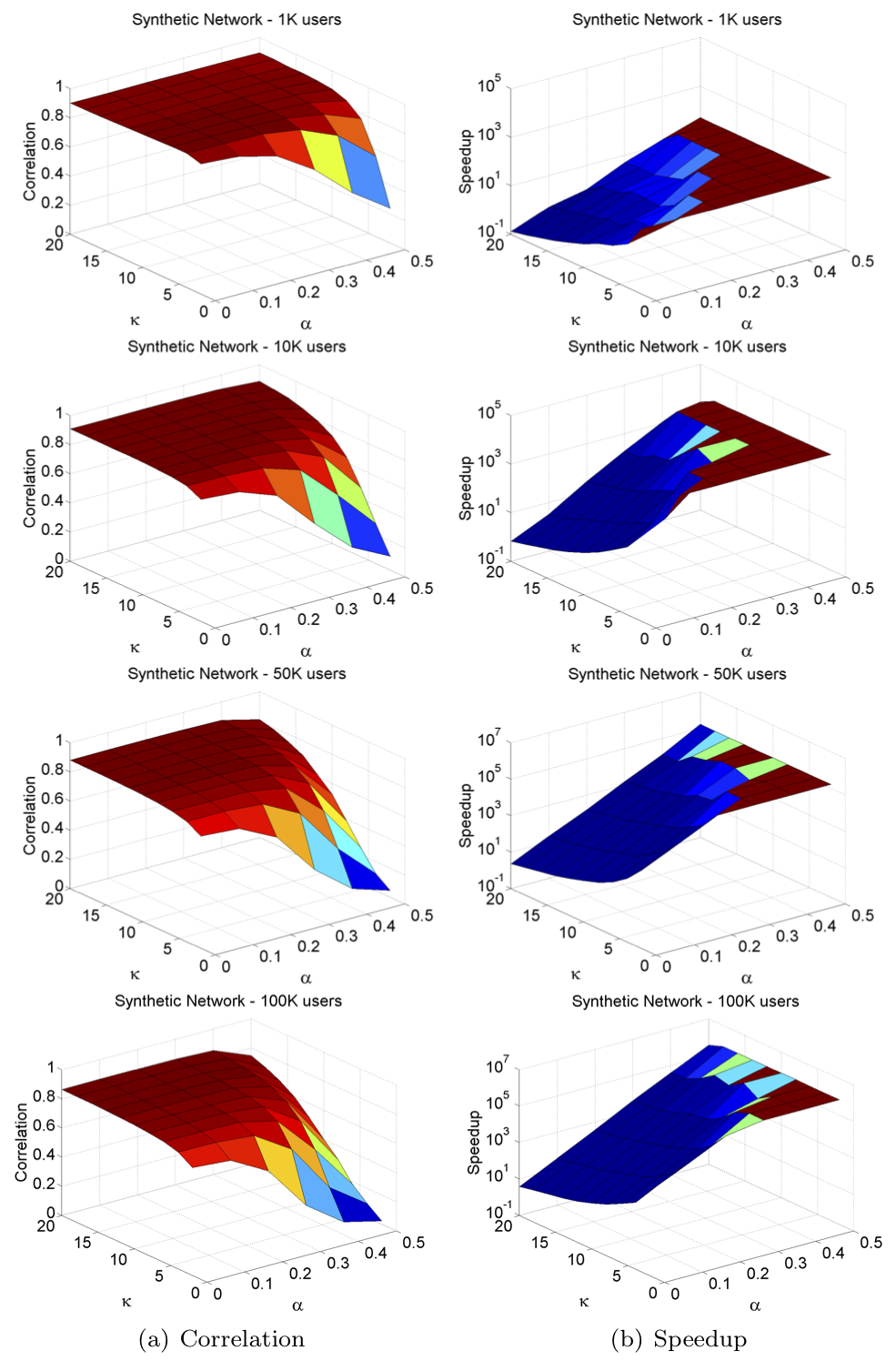}

	\end{center}
	\vspace*{-3 mm}
    	\caption{\small RA-$\kappa$path correlation (a) and speedup (b) for all the synthetic networks (size between $1K$ and $100K$ nodes).}%
	\label{fig:corr-speedup-synthetic}
\end{figure}

We computed the correlation and speedup of $\rakpath$ with
respect to Brandes' algorithm for the real and synthetic social
networks for $\kappa$ varying from $2$ to $20$ in increments of $2$
and $\alpha$ varying from $0$ to $0.5$ in increments of $0.1$.
In Figures~\ref{fig:corr-speedup-10K}--~\ref{fig:corr-speedup-100K} we present the correlation and speedup of the real networks with i) sizes below $10K$ nodes (Figures~\ref{fig:corr-speedup-10K}(a) and~\ref{fig:corr-speedup-10K}(b)), ii) sizes between $10K$ and $50K$ nodes (Figures~\ref{fig:corr-speedup-50K}(a) and~\ref{fig:corr-speedup-50K}(b)) and iii) sizes between $50K$ and $100K$ nodes (Figures~\ref{fig:corr-speedup-100K}(a) and~\ref{fig:corr-speedup-100K}(b)).
We present in Figures~\ref{fig:corr-speedup-synthetic}(a) and~\ref{fig:corr-speedup-synthetic}(b) the correlation and speedup of all synthetic networks used with sizes between $1K$ and $100K$ nodes.
These values are averages of ten executions on the ten independently generated networks for each size (thus, $10 \times 10 = 100$ runs for each network size).

We found that, as $\alpha$ decreases, 
(1) the correlation of $\rakpath$ with respect to  Brandes' algorithm increases, and 
(2) the speedup of $\rakpath$ with respect to  Brandes' algorithm decreases. 
The best correlation results are found for $\alpha = 0$ and $\kappa = 20$.
Nevertheless, for these values of $\alpha$ and $\kappa$, the runtime speedup of $\rakpath$ in comparison to Brandes' algorithm suffers the most.
Furthermore, the improvement of the correlation of $\rakpath$ across different values for $\kappa$, given a constant value of $\alpha$,
shows that the length of the path allowed to take in $\rakpath$ is extremely important to achieve better results.
The correlation performance follows a similar pattern across all network sizes and types.

In particular, we observed that for small real networks such as the first 6 networks ($<10K$ nodes),
the maximum correlation of $\rakpath$ with Brandes' algorithm is $\sim$$0.75$ to $\sim$$0.95$
and the $\rakpath$ runtime is in the order of $\sim$$30$ to $\sim$$50$ times faster than Brandes' algorithm.
For larger real networks, the maximum correlation is somewhat lower ($\sim$$0.70$ to $\sim$$0.90$).
However, the runtime of $\rakpath$ is about $\sim$$10^{2}$ to $\sim$$10^{4}$ times faster than Brandes' algorithm.
The speedup of the runtime of $\rakpath$ exhibits further improvements on the synthetic social networks,
and especially for the networks of larger size.

Overall, the maximum correlation achieved is in the range of $\sim$$0.70$ to $\sim$$0.95$ 
and the maximum speedup achieved is in the range of $\sim$$10^{2}$ to $\sim$$10^{6}$, 
depending on the values of $\alpha$, $\kappa$, and the size of the network (real or synthetic).
A general observation from these results is that we can achieve a near optimal performance of $\rakpath$ in both
correlation and speedup performance metrics when,
for a network of $n$ vertices and $m$ edges, $\alpha$ is set to $0.2$ and $\kappa$ is set to $\ln(n+m)$.
We used these values of $\alpha$ and $\kappa$ in the following experiments that compare the performance  of
$\rakpath$ with RA-Brandes and AS-Brandes.

\subsection{Comparison with RA-Brandes and AS-Brandes}
\label{sec:comparison-approx-algs}

\begin{figure}[h]
\begin{center}
\includegraphics[scale=0.5]{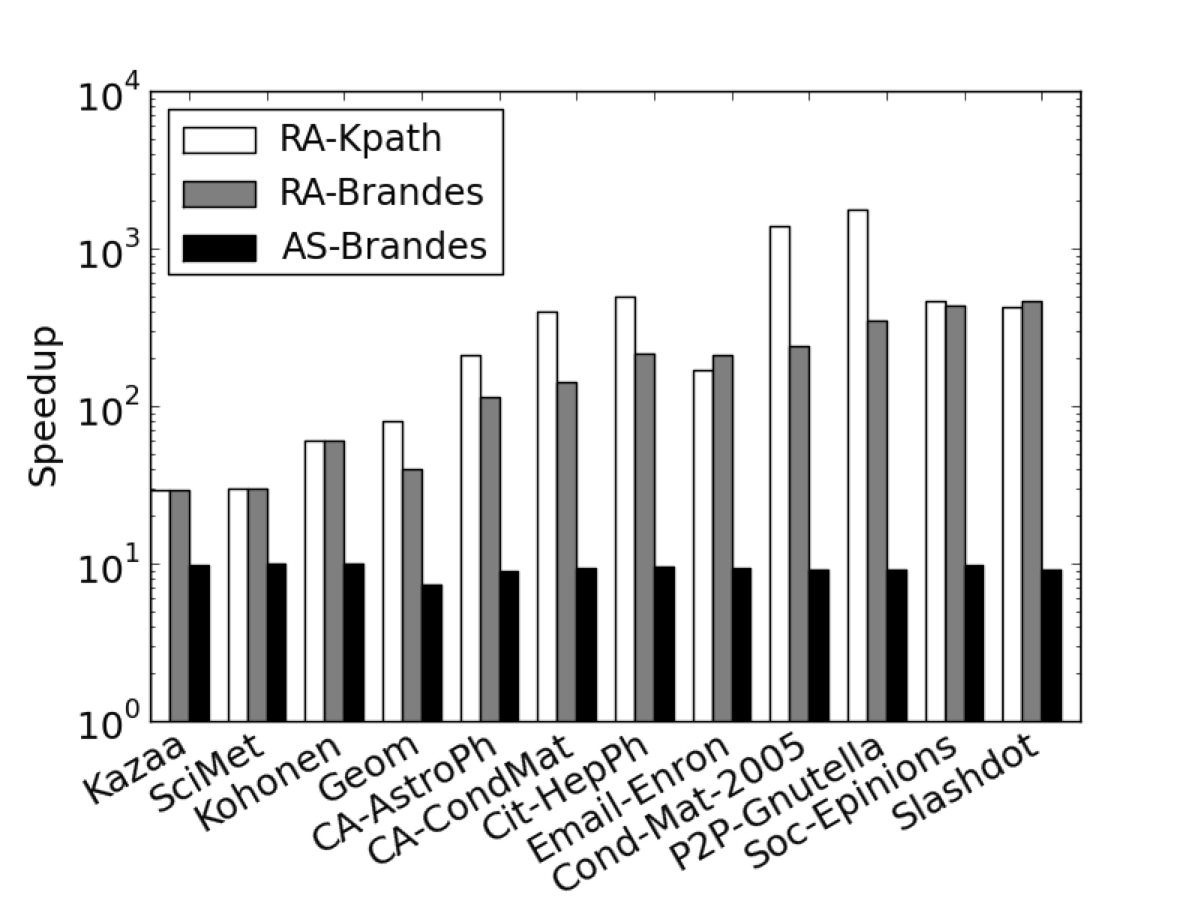}
\end{center}
\vspace*{-3 mm}
\caption{\small Speedups of RA-$\kappa$path, RA-Brandes, and AS-Brandes with respect to Brandes' algorithm for real networks.
The parameters used are $\alpha = 0.2$, $\kappa = \ln (n+m)$, $\epsilon=0.5, s=20$, and $c=5$.}
\label{fig:SpeedupComparisonRealUnmatched}
\end{figure}

\begin{figure}[h]
\begin{center}
\includegraphics[scale=0.5]{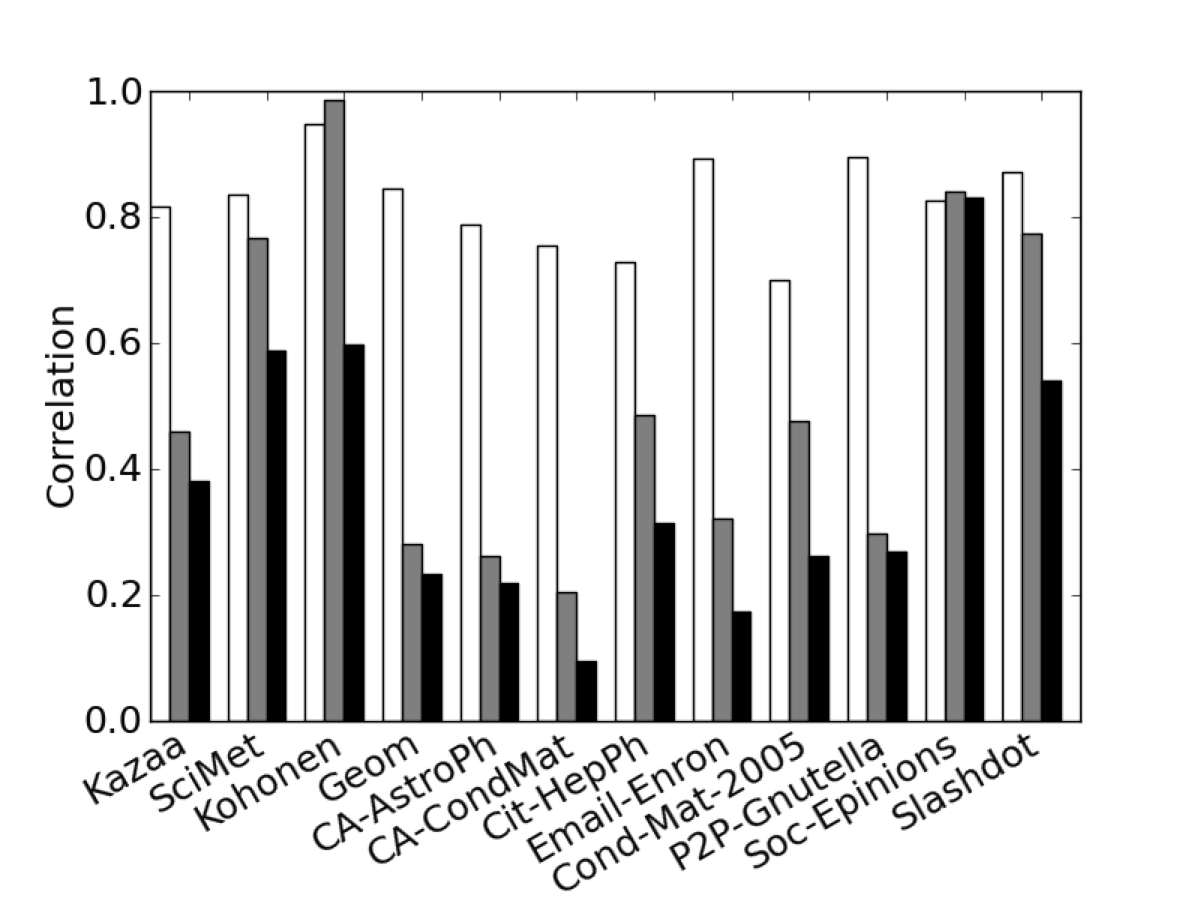}
\end{center}
\vspace*{-3 mm}
\caption{\small Correlations of RA-$\kappa$path, RA-Brandes, and AS-Brandes with respect to Brandes' algorithm for real networks.
The parameters used are $\alpha = 0.2$, $\kappa = \ln (n+m)$, $\epsilon=0.5, s=20$, and $c=5$.}
\label{fig:CorrelationComparisonRealUnmatched}
\end{figure}

Figures~\ref{fig:SpeedupComparisonRealUnmatched} and~\ref{fig:CorrelationComparisonRealUnmatched} show the correlation and speedup
results of the three algorithms ($\rakpath$, RA-Brandes, and AS-Brandes) with respect to Brandes' algorithm on real networks.
These results were obtained for $\epsilon$=$0.5$ for RA-Brandes, and $s$=$20$ and $c$=$5$ for AS-Brandes.
This choice of parameters for AS-Brandes was also used in~\cite{bad-kin-mad-mih:c:approx-betweenness}.
 The results demonstrate the superiority of $\rakpath$ over the other two algorithms in both performance metrics for most of the real networks examined.

However, we believe that the choice of parameter values $\epsilon$=$0.5$ and
$s$=$20$ is not suitable for the sizes of the networks we examined:
For example, in \cite{bad-kin-mad-mih:c:approx-betweenness}
where these values for parameters $s$ and $c$ are used in AS-Brandes,
the largest networks evaluated have $<$$10K$ nodes and $<$$50K$ edges.
For this reason, we decided to match the speedups of the three algorithms
in order to infer less biased parameter values for AS-Brandes and
RA-Brandes.  We thus performed several experiments
with various values of $\epsilon$ (for RA-Brandes) and $s$ (for
AS-Brandes), and settled on the following heuristic that helped us to 
closely match the speedups of the three algorithms with respect to Brandes' algorithm:\\
-- $\epsilon$ $=$ $2 \times ((\textnormal{RA-$\kappa$path speedup}) \times \ln (n)/n)^{1/2}$ and\\
-- $s=2 \times (\textnormal{RA-$\kappa$path speedup})$

\begin{figure}
\begin{center}
\includegraphics[scale=0.5]{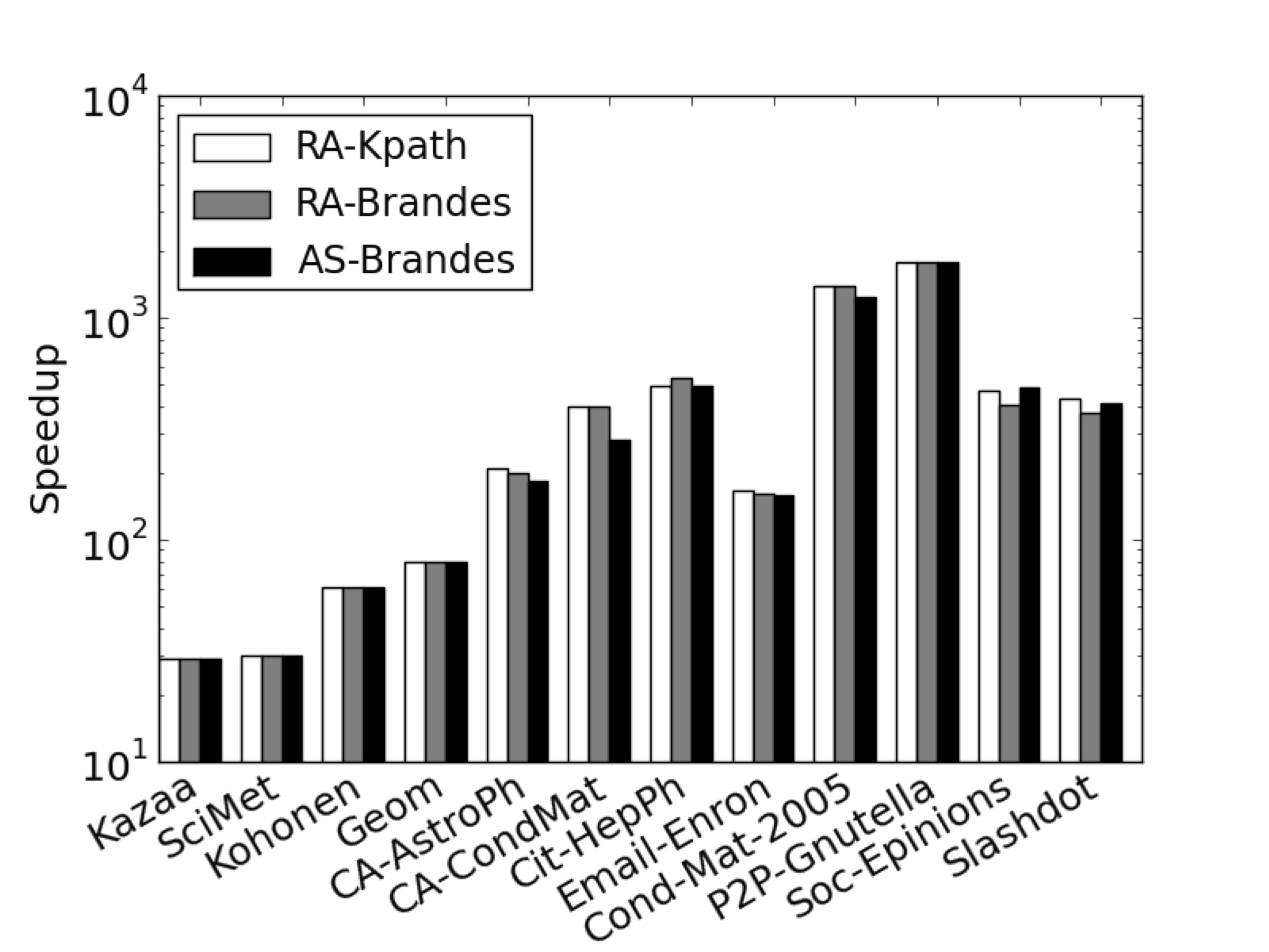}
\end{center}
\vspace*{-3 mm}
\caption{The speedups of the three algorithms on the real networks were matched to set values of their parameters for the correlation experiments.}
\label{fig:real-speedups-matched}
\end{figure}

\begin{figure}
\begin{center}
\includegraphics[scale=0.5]{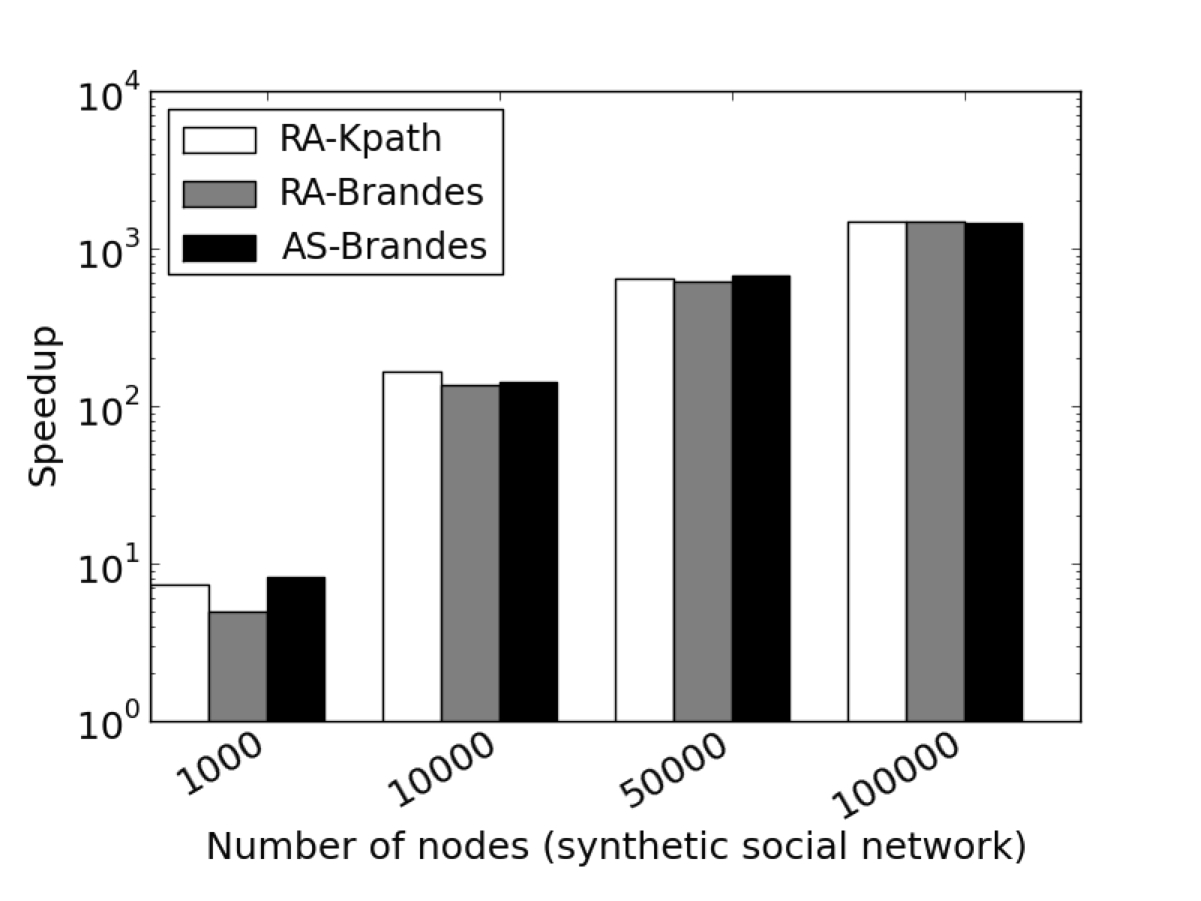}
\end{center}
\vspace*{-3 mm}
\caption{The speedups of the three algorithms on the synthetic networks were matched to set values of their parameters for the correlation experiments.}
\label{fig:synthetic-speedups-matched}
\end{figure}

The intuition for this choice of $\epsilon$ is as follows: RA-Brandes
considers dependency scores of  $\Theta((\ln n)/\epsilon^2)$ pivots
while Brandes' algorithm considers these scores of all $n$ pivots, and
so   RA-Brandes speedup can be estimated to $\Theta(n\epsilon^2/\ln
n)$;  setting this estimate to $\rakpath$ speedup yields the above
expression for $\epsilon$.  The intuition for the choice of $s$
follows a similar reasoning.
Figures~\ref{fig:real-speedups-matched} and~\ref{fig:synthetic-speedups-matched} demonstrate this process
for the real and synthetic networks respectively.
For the synthetic social graphs, the values presented are averaged
over ten executions on the ten independently generated networks for each
size (thus, $10 \times 10 = 100$ runs for each network size).

\begin{figure}
\begin{center}
\includegraphics[scale=0.5]{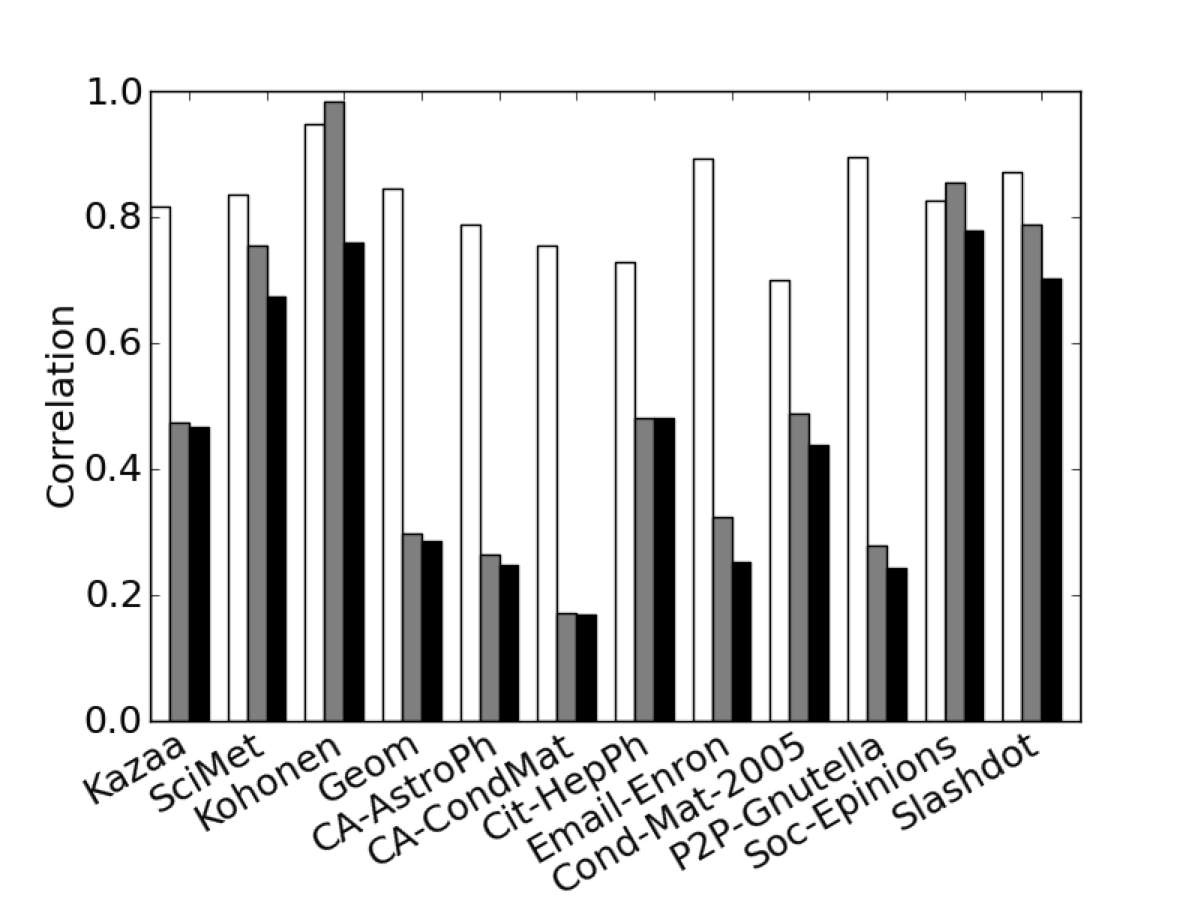}
\end{center}
\vspace*{-3 mm}
\caption{Correlations of RA-$\kappa$path, RA-Brandes, and AS-Brandes with respect to Brandes' algorithm for the real networks.
The speedups of the three algorithms were first matched to set values of their parameters and then the algorithms were ran with these values to compute their correlation scores.}
\label{fig:real-correlations-matched}
\end{figure}

\begin{figure}
\begin{center}
\includegraphics[scale=0.5]{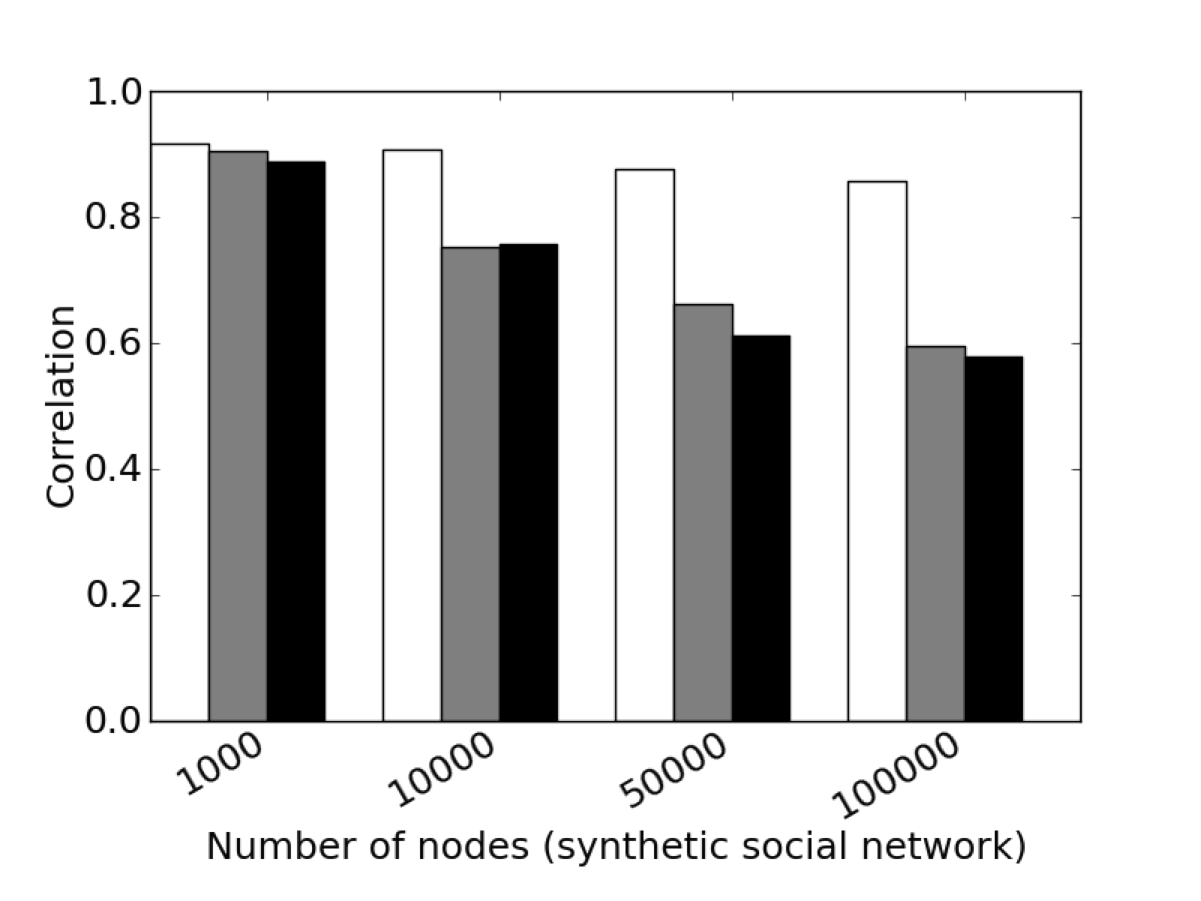}
\end{center}
\vspace*{-3 mm}
\caption{Correlations of RA-$\kappa$path, RA-Brandes, and AS-Brandes with respect to Brandes' algorithm for the synthetic networks.
The speedups of the three algorithms were first matched to set values of their parameters and then the algorithms were ran with these values to compute their correlation scores.}
\label{fig:synthetic-correlations-matched}
\end{figure}

Figures~\ref{fig:real-correlations-matched} and~\ref{fig:synthetic-correlations-matched}
show that the correlations of $\rakpath$, RA-Brandes, and AS-Brandes vary widely when their 
speedups are matched.
If we compare the results in Figure~\ref{fig:real-correlations-matched} with the previous correlation performance results shown in Figure~\ref{fig:CorrelationComparisonRealUnmatched}, we also notice that the correlations of RA-Brandes and AS-Brandes are enhanced during the speedup-matching process.

Overall, these real networks exhibit a wide range of correlation performance because they acquire different graph properties due to their diverse domains.
In most cases (except for the $Kohonen$ and $Soc$-$Epinions1$ networks), $\rakpath$ outperforms the other two algorithms by a correlation difference of $0.1$ to $0.6$, depending on the network type and size.
The synthetic networks, on the other hand, are embedded with generic graph properties of real social networks such as power-law degree distribution and high average clustering coefficient.
These networks maintain the particular graph properties while increasing the graph size and demonstrate that $\rakpath$ is consistently better than the other two algorithms on larger networks.

The better performance of $\rakpath$ shown in these results, even after we matched its speedup with the other algorithms, demonstrates that our proposed algorithm can be used to calculate more accurately
relative ranks of betweenness scores for the nodes in a network and could be ideal for experiments
on large networks where reliable results are needed fast.

Table~\ref{tab:topNreal} shows top $N\%$ $\rakpath$ (RA-K), top $N\%$ RA-Brandes (RA-B), and top $N\%$ AS-Brandes (AS-B), for the real and synthetic social networks and for $N$=$1$, $5$, and $10$. 
The results shown were obtained after the algorithms were matched in speedup, as mentioned earlier. 
Overall, $\rakpath$ outperforms the other two algorithms by a significant difference of $\sim$$10\%$ to $\sim$$40\%$, 
in identifying the $top$--$1\%$ high betweenness centrality nodes, in all the sizes and types of networks.
This result stresses the effectiveness of $\rakpath$ in identifying the nodes in a social network which rank the highest in betweenness, and doing so in a fast and efficient way.

When we examine the $top$--$5\%$ and $top$--$10\%$ of nodes, we increase accordingly the subset of nodes considered for the calculation of the high betweenness node overlap.
Intuitively, this means that any of these algorithms should perform better, as more nodes are included in the subset, thus increasing the probability to find more $top$ central nodes.
This intuition is verified for the RA-Brandes and AS-Brandes algorithms.
However, this is not the case for $\rakpath$, for which we notice an overall decrease in the performance.
This performance deterioration could be due to the arbitrary ordering of low $\kappa$-path
centrality nodes arising from closeness in their values, allowing them to enter the set of $top$ central nodes.
In the future, we plan to further examine this ordering and find ways to improve the relative ranking of nodes, thus enhancing the performance of the $\rakpath$ algorithm.

\begin{landscape}

\scriptsize{
\begin{table}
\begin{center}
\footnotesize{
\caption{\small Percentage overlap of the $Top$--$N\%$ nodes computed by the $3$ algorithms with respect to the exact betweenness centrality values.
The speedups of the $3$ algorithms were first matched to set their parameters and then executed to compute the $Top$--$N\%$ overlap.
Values in bold denote the highest in the respective ($N$-value) category.}
\label{tab:topNreal}
\begin{tabular}{| l | r || c | c | c || c | c | c || c | c | c |} \hline
Network 	&	Size	&RA-K			&RA-B 					& AS-B
				&RA-K		 	&RA-B 					& AS-B
				&RA-K			&RA-B					& AS-B\\
		&		&1\%				&1\%						&1\%
				&5\%				&5\%						&5\%
				&10\%			&10\% 					&10\%\\ \hline \hline
Kazaa & 2.4K		&\textbf{79.2}		&58.3					&58.3
				&\textbf{72.7}		&64.5					&66.9
				&72.3			&79.3 					&\textbf{79.8} \\ \hline
SciMet & 2.7K		&\textbf{85.2}		&48.1					&44.4
				&\textbf{77.9}		&66.2					&64.0
				&\textbf{76.5}		&70.2 					&69.1\\ \hline
Kohonen & 3.7K         	&\textbf{75.7}		&45.9 					&64.9 	
				&64.4			&67.6					&\textbf{69.1}	
				&60.2			&\textbf{76.7}				&74.0  \\ \hline
Geom & 6.1K           	&\textbf{68.9}		&55.7					&59.0 	
				&71.0			&\textbf{84.0} 				&83.4 					
				&72.0    			&\textbf{90.4}  				&89.9  \\ \hline
CA-AstroPh & 18.7K  &\textbf{63.1} 		&42.2 					&39.6 	
				&\textbf{68.8} 		&68.1 					&68.1						
				&74.9			&77.8    					&\textbf{78.7}  	\\ \hline
CA-CondMat & 23.1K	&\textbf{74.5} 	&48.1 					&48.9 
				&\textbf{76.6} 	&73.2 					&72.4
				&76.9		&\textbf{81.8}				&\textbf{81.8}\\ \hline
Cit-HepPh & 34.5K	&\textbf{71.3} 	&53.9 					&47.8
				&\textbf{66.1} 	&61.2 					&61.4
				&66.3		&68.9 				   	&\textbf{69.7}\\ \hline
Email-Enron & 36.7K	&75.1 		&\textbf{79.0}				&76.8
				&63.8 		&88.5 					&\textbf{89.1}
				&65.6    		&\textbf{92.7}				&\textbf{92.7}\\ \hline
Cond-Mat-2005 & 40.4K     	&66.1 		&\textbf{68.6} 				&61.4 	
				&68.2 		&\textbf{86.5} 				&85.7 					
				&70.4    		&89.4    					&\textbf{89.5} 		\\ \hline
P2P-Gnutella31 & 62.5K   	&\textbf{78.2} 	&31.0 					&26.6 	
				&\textbf{78.3} 	&50.7 					&50.1 					
				&\textbf{77.4}	&66.2					&65.0 			\\ \hline
Soc-Epinions1 & 75.9K	&\textbf{80.6} 	&70.2					&71.0
				&75.0		&\textbf{90.2}				&90.0
				&72.7		&94.8 					&\textbf{95.0} \\ \hline
Soc-Slashdot0902 & 82.2K	&\textbf{85.9} 	&67.4 				&67.7
				 	&85.2 		&\textbf{88.8} 			&88.3
 					&78.4    		&\textbf{92.1}			&92.0 \\ \hline
synth-1K & 1K             &\textbf{83.0}				&70.0	&65.0
				&\textbf{82.4}				&70.6	&69.6
				&\textbf{77.3}				&70.1	&69.7		\\ \hline
synth-10K & 10K		&\textbf{88.3}				&58.0	&58.4
				&\textbf{82.4}				&67.8	&67.8
				&\textbf{78.7}				&78.5	&78.5		\\ \hline
synth-50K & 50K 	    	&\textbf{86.6}				&61.6	&60.8
				&\textbf{81.7}				&76.5	&77.0
				&77.5					&83.5	&\textbf{83.8}		\\ \hline
synth-100K & 100K 	&\textbf{87.5}				&61.0	&60.4
				&\textbf{81.4}				&79.7	&79.8
				&77.1					&84.4	&\textbf{84.6}		\\ \hline

\end{tabular}
}
\end{center}
\end{table}
}
\normalsize

\end{landscape}

\section{Summary and Discussions}
\label{sec:summary}

In this paper, we introduced a new graph centrality index called $\kappa$-path centrality and 
presented a randomized algorithm $\rakpath$ for estimating its value
for all vertices. Our experimental evaluation demonstrates that this
centrality metric can be used to estimate in a scalable way the relative importance 
of nodes as per the betweenness centrality index: the correlation between the exact and approximate centrality indices
is between $0.70$ and $0.95$ for all network sizes for a speedup gain
of up to 6 orders of magnitude for networks with more than $10K$
nodes.

Our experiments also show that $\rakpath$ is very effective and fast in identifying
the $top$--$1\%$ or the $top$--$5\%$ nodes in the  exact betweenness score,
outperforming previously known approximate betweenness centrality
algorithms AS-Brandes~\cite{bad-kin-mad-mih:c:approx-betweenness} and RA-Brandes~\cite{bra-pic:j:central-estimate}.
The near optimal performance of $\rakpath$ in both correlation and speedup performance metrics can be achieved
when its parameters are set to $\alpha$=$0.2$ and $\kappa$=$\ln(n+m)$, for a network of $n$ number of nodes and $m$ number of edges.

Through our experiments, we have shown that $\kappa$-path centrality can be used as 
an alternative to node betweenness centrality since (a) ~$\kappa$-path
centrality closely models the spread of information  
in a network and allows to quantify the influence of any node in the
network and (b)~the speedup performance of $\rakpath$ for  
estimating $\kappa$-path centrality surpasses those achieved by existing methods of 
computing exact or approximate betweenness centrality values. 

In fact, a parallelized version of our proposed randomized $\rakpath$ algorithm
has been successfully used in a study of the Steam Community~\cite{blackburn12games-cheaters}, 
a large-scale gaming social network with over $12$ million players and $88.5$ million social edges.
Our randomized algorithm was used to approximate the betweenness centrality of players and help
identify top central players in the gaming social network.

There are various practical applications for identifying the top betweenness centrality nodes in large networks.
For example, in unstructured peer-to-peer overlays the high betweenness peers have a
significant impact since they relay much of the traffic~\cite{kourtellis11p2pcentrality}.
If under-provisioned, they can damage the overall system performance.
If malicious, they can snoop on or divert significant communication.
Alternatively, they are great monitoring locations for examining the network communication for traffic characterization studies.

Therefore, identifying the top betweenness centrality nodes can have impact on the network performance (through resource
provisioning), security (by restricting the monitoring of potential
malicious activity to a small group of candidates), and traffic characterization.
Deterministically identifying high betweenness nodes in such a network is infeasible not only because of the large scale
(typically hundreds of thousands or millions of nodes) but also because of their dynamic nature caused by high node churn.

Another example of the applicability of our approach is efficient data placement and diffusion.
For example, data can be placed on a few high betweenness centrality nodes in a large communication network,
such as the web graph, where informed data placement may lead to faster access to event announcements,
or a mobile phones network, where data can be security patches that can be efficiently propagated from
a few targeted central individuals to the rest of the population.

\section*{Acknowledgment}
This research was partially supported by the
National Science Foundation under Grants No. CNS-0831785 and
CNS-0952420. The authors 
would also like to acknowledge the use of the computing services provided by Research
Computing, University of South Florida.  

\bibliographystyle{abbrv}
\bibliography{refs}

\end{document}